\documentclass[twocolumn,aps,prc,superscriptaddress,showpacs,floatfix]{revtex4}
\usepackage{amssymb}
\usepackage{amsmath}
\usepackage{graphicx}
\usepackage{longtable}
\usepackage[normalem]{ulem}
\usepackage[dvips]{color}

\renewcommand\sout{\bgroup \color{red} \ULdepth=-.5ex \ULset}

\begin{document}

\title{Nuclear matter fourth-order symmetry energy in the relativistic
mean field model}
\author{Bao-Jun Cai}
\affiliation{Department of Physics, Shanghai Jiao Tong University, Shanghai 200240, China}
\author{Lie-Wen Chen\footnote{%
Corresponding author (email: lwchen$@$sjtu.edu.cn)}}
\affiliation{Department of Physics, Shanghai Jiao Tong University, Shanghai 200240, China}
\affiliation{Center of Theoretical Nuclear Physics, National Laboratory of Heavy Ion
Accelerator, Lanzhou 730000, China}
\date{\today}

\begin{abstract}
Within the nonlinear relativistic mean field model, we derive the analytical
expression of the nuclear matter fourth-order symmetry energy $E_{\mathrm{{%
sym},4}}(\rho )$. Based on two accurately calibrated interactions
FSUGold and IU-FSU, our results show that the value of
$E_{\mathrm{{sym},4}}(\rho )$ at normal nuclear matter density $\rho
_{0}$ is generally less than $1$ MeV, confirming the empirical
parabolic approximation to the equation of state for asymmetric
nuclear matter at $\rho _{0}$. On the other hand, we find that the
$E_{\mathrm{{sym},4}}(\rho )$ may become nonnegligible at high
densities. Furthermore, the analytical form of the $E_{\mathrm{{sym},4}%
}(\rho )$ provides the possibility to study the higher-order effects on the
isobaric incompressibility of asymmetric nuclear matter, i.e., $K_{\mathrm{%
sat}}(\delta )=K_{0}+K_{\mathrm{{sat},2}}\delta ^{2}+K_{\mathrm{{sat},4}%
}\delta ^{4}+\mathcal{O}(\delta ^{6})$ where $\delta =(\rho _{n}-\rho
_{p})/\rho $ is the isospin asymmetry, and we find that the value of $K_{%
\mathrm{{sat},4}}$ is generally small compared with that of the $K_{\mathrm{{%
sat},2}}$. In addition, we study the effects of the $E_{\mathrm{{sym},4}%
}(\rho )$ on the proton fraction $x_{p}$ and the core-crust transition
density $\rho _{t}$ and pressure $P_{t}$ in neutron stars. Interestingly, we
find that, compared with the results from the empirical parabolic
approximation, including the $E_{\mathrm{{sym},4}}(\rho )$ contribution can
significantly enhance the $x_{p}$ at high densities and strongly reduce the $%
\rho _{t}$ and $P_{t}$ in neutron stars, demonstrating that the widely used
empirical parabolic approximation may cause large errors in determining the $%
x_{p}$ at high densities as well as the $\rho _{t}$ and $P_{t}$ in neutron
stars within the nonlinear relativistic mean field model, consistent with
previous nonrelativistic calculations.
\end{abstract}

\maketitle

\section{Introduction}

\label{s1}

One of fundamental issues in nuclear physics is the equation of
state (EOS) of isospin asymmetric nuclear matter, which plays a
central role in understanding not only the structure of radioactive
nuclei, the reaction dynamics induced by rare isotopes, and the
liquid-gas phase transition in asymmetric nuclear matter, but also
many critical issues in astrophysics
\cite{LiBA98,danie02,lattimer04,baran05,steiner05,chen07,li08}. For
symmetric nuclear matter with equal fractions of neutrons and
protons, its EOS is relatively well-determined from analyses of the
giant monopole resonances of finite nuclei \cite{youngblood99,LiT07}
as well as collective flows \cite{danie02} and subthreshold kaon
production \cite{aic85, fuchs06} in relativistic nucleus-nucleus
collisions. On the other hand, the EOS of asymmetric nuclear matter,
especially the density dependence of the nuclear symmetry energy
$E_{\mathrm{sym}}(\rho )$, is poorly known. During the last decade,
significant progress has been made both experimentally and
theoretically on constraining the behavior of the symmetry energy
around and below normal nuclear matter
density~\cite{Che05,Tsa09,Cen09,Nat10} (See, e.g., Refs.
\cite{XuC10,Che10,Tsa11,Che11a,New11} for review of recent progress)
while its super-normal density behavior remains elusive and largely
controversial~\cite{Xia09,Fen10,Rus11,XuC10b}. Theoretically, all
many-body theory calculations to date have demonstrated that the
nuclear symmetry energy essentially characterizes the isospin
dependent part of the EOS of asymmetric nuclear matter and the
higher-order terms in isospin
asymmetry are unimportant, at least for densities up to moderate values \cite%
{li08}, leading to the well-known empirical parabolic law.

When the empirical parabolic law itself provides a good approximation to the
EOS of asymmetric nuclear matter and thus allows one to extract the symmetry
energy from the energy difference between pure neutron matter and symmetric
nuclear matter, it may cause large errors when it is applied to determine
some physical quantities under special conditions. For example, the
higher-order terms in isospin asymmetry presented in the EOS of asymmetric
nuclear matter at supra-normal densities can significantly modify the proton
fraction in $\beta $-equilibrium neutron-star matter and the critical
density for the direct Urca process which can lead to faster cooling of
neutron stars \cite{Zha01,steiner08}. In addition, recent studies \cite{xu09}
indicate that the higher-order terms in isospin asymmetry are very important
for determining the transition density and pressure at the inner edge
separating the liquid core from the solid crust of neutron stars where the
matter is extremely neutron-rich. Furthermore, the higher-order effects on
the incompressibility of asymmetric nuclear matter have also been studied
recently \cite{Che09}. These studies about the higher-order effects are
essentially performed within the nonrelativistic models since the analytical
expressions of the higher-order terms in isospin asymmetry, e.g., the
nuclear matter fourth-order symmetry energy $E_{\mathrm{{sym},4}}(\rho )$,
can be relatively easily obtained in such nonrelativistic models. It is thus
interesting to see if the same conclusion can be obtained within the
relativistic models.

One of very popular relativistic models is the relativistic mean field (RMF)
model which is generally based on effective interaction Lagrangians
involving nucleon and meson fields \cite{Wal74}. As a phenomenological
approach, the RMF model has achieved great success during the last decades
in describing many nuclear phenomena \cite{Ser86,Men06}. Although the full
expressions have been usually used in realistic RMF model calculations,
nevertheless, it will be instructive to see separately the effects of the
higher-order terms in isospin asymmetry, e.g., the nuclear matter
fourth-order symmetry energy $E_{\mathrm{{sym},4}}(\rho )$, within the RMF
model. The main motivation of the present work is to derive the nuclear
matter fourth-order symmetry energy $E_{\mathrm{{sym},4}}(\rho )$ within the
nonlinear RMF model and then explore the higher-order $E_{\mathrm{{sym},4}%
}(\rho )$ corrections to the widely used empirical parabolic law for the
isospin asymmetric nuclear matter. Based on two accurately calibrated
interactions, our results indicate that the $E_{\mathrm{{sym},4}}(\rho )$
may have significant influence on the properties of isospin asymmetric
nuclear matter, the proton fraction $x_{p}$ in $\beta $-stable neutron star
matter and the core-crust transition density $\rho _{t}$ and pressure $P_{t}$
in neutron stars, confirming the previous nonrelativistic calculations.

The paper is organized as follows. In Section \ref{s2}, we briefly discuss
the bulk characteristic parameters of asymmetric nuclear matter. The model
used in the present paper and the analytical expression of the $4$th-order
symmetry energy will also given in this section. The results and discussions
are then presented in Section \ref{s3}. Finally, a summary is given in
Section \ref{s4}.

\section{Theoretical formulism}

\label{s2}

\subsection{Characteristic parameters of asymmetric nuclear matter}

The EOS of isospin asymmetric nuclear matter, defined by its binding energy
per nucleon, can be expanded to $4$nd-order in isospin asymmetry $\delta $ as%
\begin{equation}
E(\rho ,\delta )=E_{0}(\rho )+E_{\mathrm{sym}}(\rho )\delta ^{2}+E_{\mathrm{{%
sym},4}}(\rho )\delta ^{4}+\mathcal{O}(\delta ^{6}),  \label{EOSANM}
\end{equation}%
where $\rho =\rho _{n}+\rho _{p}$ is the baryon density with $\rho _{n}$ and
$\rho _{p}$ denoting the neutron and proton densities, respectively; $\delta
=(\rho _{n}-\rho _{p})/(\rho _{p}+\rho _{n})$ is the isospin asymmetry; $%
E_{0}(\rho )=E(\rho ,\delta =0)$ is the binding energy per nucleon in
symmetric nuclear matter; the nuclear matter symmetry energy $E_{\mathrm{sym}%
}(\rho )$ and the $4$th-order symmetry energy $E_{\mathrm{{sym},4}}(\rho )$
are expressed, respectively, as
\begin{eqnarray}
E_{\mathrm{sym}}(\rho ) &=&\left. \frac{1}{2!}\frac{\partial ^{2}E(\rho
,\delta )}{\partial \delta ^{2}}\right\vert _{\delta =0},~~  \label{Esym} \\
E_{\mathrm{{sym},4}}(\rho ) &=&\left. \frac{1}{4!}\frac{\partial ^{4}E(\rho
,\delta )}{\partial \delta ^{4}}\right\vert _{\delta =0}.  \label{Esym4}
\end{eqnarray}%
In Eq. (\ref{EOSANM}), the absence of odd-order terms in $\delta $ is due to
the exchange symmetry between protons and neutrons in nuclear matter when
one neglects the Coulomb interaction and assumes the charge symmetry of
nuclear forces. The higher-order (including the $4$th-order) coefficients in
$\delta $ are usually very small. For example, the magnitude of the $\delta
^{4}$ term at normal nuclear matter density $\rho _{0}$ is estimated to be
less than $1$ MeV in microscopic many-body approaches \cite{sie70,sj74,lag81}
and also in phenomenological nonrelativistic models \cite{Che09} as well as
relativistic models as will be shown in this work. Neglecting the
contribution from higher-order terms in Eq. (\ref{EOSANM}) leads to the
well-known empirical parabolic law, i.e., $E(\rho ,\delta )\simeq E_{0}(\rho
)+E_{\mathrm{sym}}(\rho )\delta ^{2}$ for the EOS of asymmetric nuclear
matter and the symmetry energy $E_{\mathrm{sym}}(\rho )$ can thus be
extracted from $E_{\mathrm{sym}}(\rho )\simeq E(\rho ,\delta =1)-E(\rho
,\delta =0)$.

Around normal nuclear matter density $\rho _{0}$, the $E_{0}(\rho )$ can be
expanded, e.g., up to $4$th-order in density, as,
\begin{equation}
E_{0}(\rho )=E_{0}(\rho _{0})+\frac{K_{0}}{2!}\chi ^{2}+\frac{J_{0}}{3!}\chi
^{3}+\frac{I_{0}}{4!}\chi ^{4}+\mathcal{O}(\chi ^{5}),  \label{DenExp0}
\end{equation}%
where $\chi $ is a dimensionless variable characterizing the deviations of
the density from normal nuclear matter density $\rho _{0}$ and it is
conventionally defined as
\begin{equation}
\chi =\frac{\rho -\rho _{0}}{3\rho _{0}}.  \label{DefTh}
\end{equation}%
The first term $E_{0}(\rho _{0})$ on the right-hand-side (r.h.s) of Eq. (\ref%
{DenExp0}) is the binding energy per nucleon in symmetric nuclear matter at
normal nuclear matter density $\rho _{0}$ and the coefficients of other
terms are,
\begin{eqnarray}
K_{0} &=&\left. 9\rho _{0}^{2}\frac{\partial ^{2}E_{0}(\rho )}{\partial \rho
^{2}}\right\vert _{\rho =\rho _{0}},~~  \label{K0} \\
J_{0} &=&\left. 27\rho _{0}^{3}\frac{\partial ^{3}E_{0}(\rho )}{\partial
\rho ^{3}}\right\vert _{\rho =\rho _{0}},  \label{J0} \\
I_{0} &=&\left. 81\rho _{0}^{4}\frac{\partial ^{2}E_{0}(\rho )}{\partial
\rho ^{4}}\right\vert _{\rho =\rho _{0}}.  \label{I0}
\end{eqnarray}%
The linear $\chi $ term on the r.h.s of Eq. (\ref{DenExp0}) vanishes
according to the definition of the saturation density $\rho _{0}$. The
coefficient $K_{0}$ is the well-known incompressibility coefficient of
symmetric nuclear matter and it characterizes the curvature of $E_{0}(\rho
_{0})$ at $\rho _{0}$. The coefficients $J_{0}$ and $I_{0}$ are the $3$%
rd-order and $4$th-order incompressibility coefficients of symmetric nuclear
matter \cite{Che09}, respectively.

Similarly, around normal nuclear matter density $\rho _{0}$, the symmetry
energy $E_{\mathrm{sym}}(\rho )$ and the $4$th-order symmetry energy $E_{%
\mathrm{{sym},4}}(\rho )$ can be expanded, e.g., up to $4$th-order in $\chi $%
, as,%
\begin{eqnarray}
E_{\mathrm{sym}}(\rho ) &=&E_{\mathrm{sym}}(\rho _{0})+L\chi +\frac{K_{%
\mathrm{sym}}}{2!}\chi ^{2}  \notag \\
&&+\frac{J_{\mathrm{sym}}}{3!}\chi ^{3}+\frac{I_{\mathrm{sym}}}{4!}\chi ^{4}+%
\mathcal{O}(\chi ^{5}),  \label{DenExp2}
\end{eqnarray}%
and
\begin{eqnarray}
E_{\mathrm{{sym},4}}(\rho ) &=&E_{\mathrm{{sym},4}}(\rho _{0})+L_{\mathrm{{%
sym},4}}\chi +\frac{K_{\mathrm{{sym},4}}}{2!}\chi ^{2}  \notag \\
&&+\frac{J_{\mathrm{{sym},4}}}{3!}\chi ^{3}+\frac{I_{\mathrm{{sym},4}}}{4!}%
\chi ^{4}+\mathcal{O}(\chi ^{5}),  \label{DenExp4}
\end{eqnarray}%
respectively, where the $L$, $K_{\mathrm{sym}}$, $J_{\mathrm{sym}}$, $I_{%
\mathrm{sym}}$ and $L_{\mathrm{{sym},4}}$, $K_{\mathrm{{sym},4}}$, $J_{%
\mathrm{{sym},4}}$, $I_{\mathrm{{sym},4}}$ are the slope parameter,
curvature parameter, $3$rd-order and $4$th-order density coefficients of the
$E_{\mathrm{sym}}(\rho )$ and $E_{\mathrm{{sym},4}}(\rho )$ at $\rho _{0}$,
respectively, whose definitions are similar to Eq. (\ref{K0}) - Eq. (\ref{I0}%
). In general, these characteristic parameters can be written as,
\begin{equation}
W_{ij}=\left. (3\rho _{0})^{j}\frac{\partial ^{j}E_{\mathrm{sym},2i}(\rho )}{%
\partial \rho ^{j}}\right\vert _{\rho =\rho _{0}},~~i,j=1,2,\cdots
\label{CoExp}
\end{equation}%
for example, $W_{11}=L$, $W_{12}=K_{\mathrm{sym}}$, $W_{23}=J_{\mathrm{{sym}%
,4}}$, $W_{24}=I_{\mathrm{{sym},4}}$, and so on.

In the above Taylor's expansions, we have kept all terms up to $4$th-order
in $\delta $ or $\chi $. The $14$ characteristic parameters, namely, $%
E_{0}(\rho _{0})$, $K_{0}$, $J_{0}$, $I_{0}$, $E_{\mathrm{sym}}(\rho _{0})$,
$L$, $K_{\mathrm{sym}}$, $J_{\mathrm{sym}}$, $I_{\mathrm{sym}}$, $E_{\mathrm{%
{sym},4}}(\rho _{0})$, $L_{\mathrm{{sym},4}}$, $K_{\mathrm{{sym},4}}$, $J_{%
\mathrm{{sym},4}}$ and $I_{\mathrm{{sym},4}}$ are well-defined, and they
characterize the EOS of an asymmetric nuclear matter and its density
dependence at normal nuclear matter density $\rho _{0}$. Among these
parameters, $E_{0}(\rho _{0})$, $K_{0}$, $E_{\mathrm{sym}}(\rho _{0})$, $L$
and $K_{\mathrm{sym}}$ have been extensively studied in the literature and
significant progress has been made over past few decades \cite{li08}.

The incompressibility of asymmetric nuclear matter is an important quantity
to characterize its EOS. Conventionally, the incompressibility coefficient
is defined at the saturation density where the pressure $P(\rho ,\delta
)=\rho ^{2}{\partial E(\rho ,\delta )}/{\partial \rho }=0$, and it is called
the isobaric incompressibility coefficient \cite{prak85} given by%
\begin{equation}
K_{\mathrm{sat}}(\delta )=\left. 9\rho _{\mathrm{sat}}^{2}\frac{\partial
^{2}E(\rho ,\delta )}{\partial \rho ^{2}}\right\vert _{\rho =\rho _{\mathrm{%
sat}}}.  \label{isoincom}
\end{equation}%
The isobaric incompressibility coefficient $K_{\mathrm{sat}}(\delta )$ thus
only depends on the isospin asymmetry $\delta $. One can show that up to $4$%
th-order in $\delta $, the $K_{\mathrm{sat}}(\delta )$ can be expressed as
\cite{Che09}
\begin{equation}
K_{\mathrm{sat}}(\delta )=K_{0}+K_{\mathrm{sat,2}}\delta ^{2}+K_{\mathrm{%
sat,4}}\delta ^{4}+\mathcal{O}(\delta ^{6}),  \label{SatIncom}
\end{equation}%
with%
\begin{align}
K_{\mathrm{{sat},2}}=& K_{\mathrm{sym}}-6L-\frac{J_{0}L}{K_{0}},
\label{SatIncom2} \\
K_{\mathrm{sat,4}}=& K_{\mathrm{sym,4}}-6L_{\mathrm{sym,4}}-\frac{J_{0}L_{%
\mathrm{sym,4}}}{K_{0}}+\frac{9L^{2}}{K_{0}}-\frac{J_{\mathrm{sym}}L}{K_{0}}
\notag \\
& +\frac{I_{0}L^{2}}{2K_{0}^{2}}+\frac{J_{0}K_{\mathrm{sym}}L}{K_{0}^{2}}+%
\frac{3J_{0}L^{2}}{K_{0}^{2}}-\frac{J_{0}^{2}L^{2}}{2K_{0}^{3}}.
\label{Ksat4}
\end{align}%
If we use the parabolic approximation for the EOS of symmetric nuclear
matter, i.e., $E_{0}(\rho )=E_{0}(\rho _{0})+\frac{1}{2}K_{0}^{2}\chi +%
\mathcal{O}(\chi ^{3}),$ then the $K_{\mathrm{{sat},2}}$ parameter is
reduced to%
\begin{equation}
K_{\mathrm{asy}}=K_{\mathrm{sym}}-6L,  \label{DefKasy}
\end{equation}%
and this expression has been extensively used in the literature to
characterize the isospin dependence of the incompressibility of asymmetric
nuclear matter in the literature \cite{baran05,Che05,Cen09,LiT07,lopez88}.

\subsection{The $4$th-order symmetry energy in the nonlinear RMF model}

In the present work, we use the interacting Lagrangian density of the
nonlinear RMF model supplemented with couplings between the isoscalar and
the isovector mesons \cite{Mul96,Hor01,Tod05}, i.e.,
\begin{align}
& \mathcal{L}(\psi ,\sigma ,\omega _{\mu },\vec{\mkern1mu\rho }_{\mu })=\bar{%
\psi}\left[ \gamma _{\mu }(i\partial ^{\mu }-g_{\omega }\omega ^{\mu
})-(M-g_{\sigma }\sigma )\right] \psi \null  \notag \\
& +\frac{1}{2}\left( \partial _{\mu }\sigma \partial ^{\mu }\sigma
-m_{\sigma }^{2}\sigma ^{2}\right) -\frac{1}{3}b_{\sigma }M(g_{\sigma
}\sigma )^{3}  \notag \\
& -\frac{1}{4}c_{\sigma }(g_{\sigma }\sigma )^{4}\null+\frac{1}{2}m_{\omega
}^{2}\omega _{\mu }\omega ^{\mu }-\frac{1}{4}F_{\mu \nu }F^{\mu \nu }  \notag
\\
& +\frac{1}{4}c_{\omega }(g_{\omega }^{2}\omega _{\mu }\omega ^{\mu })^{2}%
\null+\frac{1}{2}m_{\rho }^{2}\vec{\mkern1mu\rho }_{\mu }\cdot \vec{\mkern%
1mu\rho }^{\mu }  \notag \\
& -\frac{1}{4}\vec{\mkern1muG}_{\mu \nu }\cdot \vec{\mkern1muG}^{\mu \nu
}-g_{\rho }\vec{\mkern1mu\rho }_{\mu }\cdot \bar{\psi}\gamma ^{\mu }\vec{%
\mkern1mu\tau }\psi \null  \notag \\
& +\frac{1}{2}g_{\rho }^{2}\vec{\mkern1mu\rho }_{\mu }\cdot \vec{\mkern%
1mu\rho }^{\mu }\left[ \Lambda _{S}g_{\sigma }^{2}\sigma ^{2}+\Lambda
_{V}g_{\omega }^{2}\omega _{\mu }\omega ^{\mu }\right]  \label{NonRMF}
\end{align}%
where $F_{\mu \nu }\equiv \partial _{\mu }\omega _{\nu }-\partial _{\nu
}\omega _{\mu }$ and$~\vec{\mkern1muG}_{\mu \nu }\equiv \partial _{\mu }\vec{%
\mkern1mu\rho }_{\nu }-\partial _{\nu }\vec{\mkern1mu\rho }_{\mu }$ are
strength tensors for $\omega $ field and $\rho $ field, respectively. $\psi $%
, $\sigma $, $\omega _{\mu }$, $\vec{\mkern1mu\rho }_{\mu }$ are nucleon
field, isoscalar-scalar field, isoscalar-vector field and isovector-vector
field, respectively, and the arrows denote the vector in isospin space. The $%
\Lambda _{S}$ and $\Lambda _{V}$ represent coupling constants between the
isovector $\rho $ meson and the isoscalar $\sigma $ and $\omega $ mesons,
respectively, which are important for the description of the density
dependence of the symmetry energy. In addition, $M$ is the nucleon mass and $%
m_{\sigma }$, $m_{\omega }$, $m_{\rho }$ are masses of mesons.

In the mean field approximation, after neglecting effects of fluctuation and
correlation, meson fields are replaced by their expectation values, i.e., $%
\bar{\sigma}\rightarrow \sigma $, $\bar{\omega}_{0}\rightarrow \omega _{\mu
} $, $\bar{\rho}_{0}^{(3)}\rightarrow \vec{\mkern1mu\rho }_{\mu }$, where
subscript \textquotedblleft $0$" indicates zeroth component of the
four-vector, superscript \textquotedblleft ($3$)" indicates third component
of the isospin, Furthermore, we also use in this work the non-sea
approximation which neglects the effect due to negative energy states in the
Dirac sea. The mean field equations are then expressed as 
\begin{align}
m_{\sigma }^{2}\bar{\sigma}=& g_{\sigma }\left[ \rho _{S}-b_{\sigma }M\left(
g_{\sigma }\bar{\sigma}\right) ^{2}-c_{\sigma }\left( g_{\sigma }\bar{\sigma}%
\right) ^{3}+\left( g_{\rho }\bar{\rho}_{0}^{(3)}\right) ^{2}\Lambda
_{S}g_{\sigma }\bar{\sigma}\right] \\
m_{\omega }^{2}\bar{\omega}_{0}=& g_{\omega }\left[ \rho -c_{\omega }\left(
g_{\omega }\bar{\omega}_{0}\right) ^{3}-\Lambda _{V}g_{\omega }\bar{\omega}%
_{0}\left( g_{\rho }\bar{\rho}_{0}^{(3)}\right) ^{2}\right] \\
m_{\rho }^{2}\bar{\rho}_{0}^{(3)}=& g_{\rho }\left[ \rho _{p}-\rho
_{n}-\Lambda _{S}g_{\rho }\bar{\rho}_{0}^{(3)}\left( g_{\sigma }\bar{\sigma}%
\right) ^{2}-\Lambda _{V}g_{\rho }\bar{\rho}_{0}^{(3)}\left( g_{\omega }\bar{%
\omega}_{0}\right) ^{2}\right]
\end{align}%
%
%
where
\begin{equation}
\rho =\langle \bar{\psi}\gamma ^{0}\psi \rangle =\rho _{n}+\rho _{p},~~\rho
_{S}=\langle \bar{\psi}\psi \rangle =\rho _{Sn}+\rho _{Sp},  \label{density}
\end{equation}%
are the baryon density and scalar density, respectively, with the latter
given by
\begin{eqnarray}
\rho _{SJ} &=&\frac{2}{(2\pi )^{3}}\int_{0}^{k_{F}^{J}}d\vec{k}\frac{%
M_{0}^{\ast }}{\sqrt{|\vec{k}|^{2}+{M_{0}^{\ast }}^{2}}}  \notag \\
&=&\frac{M_{J}^{\ast }}{2\pi ^{2}}\left[ k_{F}^{J}E_{F}^{J\ast }-{%
M_{0}^{\ast 2}}\ln \frac{k_{F}^{J}+E_{F}^{J\ast }}{M_{0}^{\ast }}\right]
,J=p,n.  \label{ScaDen}
\end{eqnarray}%
In the above expression, we have $E_{F}^{J\ast }=\sqrt{k_{F}^{J2}+M_{0}^{%
\ast 2}}$ with $M_{0}^{\ast }=M-g_{\sigma }\bar{\sigma}$ being the nucleon
Dirac mass and the Fermi momentum $k_{F}^{J}=k_{F}(1+\tau _{3}\delta )^{1/3}$
with $\tau _{3}=1$ for neutrons and $\tau _{3}=-1$ for protons, and $%
k_{F}=(3\pi ^{2}\rho /2)^{1/3}$ being the Fermi momentum for symmetric
nuclear matter.

The energy-momentum density tensor for the interacting Lagrangian density (%
\ref{NonRMF}) can be written as
\begin{equation}
\mathcal{T}^{\mu \nu }=\bar{\psi}i\gamma ^{\mu }\partial ^{\nu }\psi
+\partial ^{\mu }\sigma \partial ^{\nu }\sigma -F^{\mu \eta }\partial ^{\nu
}\omega _{\eta }-\vec{\mkern1muG}^{\mu \eta }\partial ^{\nu }\vec{\mkern%
1mu\rho }_{\eta }-\mathcal{L}g^{\mu \nu },  \label{EnMomTen}
\end{equation}%
where $g_{\mu \nu }=(+,-,-,-)$ is the Minkovski metric. In the mean field
approximation, the mean value of time (zero) component of the
energy-momentum density tensor is the energy density of the nuclear matter
system, i.e.,%
\begin{align}
\varepsilon =& \langle \mathcal{T}^{00}\rangle =\varepsilon _{\mathrm{kin}%
}^{n}+\varepsilon _{\mathrm{kin}}^{p}+\frac{1}{2}\left[ m_{\sigma }^{2}\bar{%
\sigma}^{2}+m_{\omega }^{2}\bar{\omega}_{0}^{2}+m_{\rho }^{2}\left( \bar{\rho%
}_{0}^{(3)}\right) ^{2}\right]  \notag \\
& +\frac{1}{3}b_{\sigma }(g_{\sigma }\bar{\sigma})^{3}+\frac{1}{4}c_{\sigma
}(g_{\sigma }\bar{\sigma})^{4}+\frac{3}{4}c_{\omega }(g_{\omega }\bar{\omega}%
_{0})^{4}  \notag \\
& +\frac{1}{2}\left( g_{\rho }\bar{\rho}_{0}^{(3)}\right) ^{2}\left[ \Lambda
_{S}(g_{\sigma }\bar{\sigma})^{2}+3\Lambda _{V}(g_{\omega }\bar{\omega}%
_{0})^{2}\right] ,
\end{align}%
where%
\begin{eqnarray}
\varepsilon _{\mathrm{kin}}^{J} &=&\frac{2}{(2\pi )^{3}}\int_{0}^{k_{F}^{J}}d%
\vec{k}\sqrt{|\vec{k}|^{2}+{M_{J}^{\ast 2}}}  \notag \\
&=&\frac{1}{\pi ^{2}}\int_{0}^{k_{F}^{J}}k^{2}dk\sqrt{k^{2}+{M_{0}^{\ast 2}}}
\notag \\
&=&\frac{1}{4}\left[ 3E_{F}^{J\ast }\rho _{J}+M_{0}^{\ast }\rho _{SJ}\right]
,~~J=p,n,  \label{EnDenKin}
\end{eqnarray}%
is the kinetic part of the energy density. Similarly, the mean value of
space components of the energy-momentum density tensor corresponds to the
pressure of the system, i.e.,%
\begin{align}
P=& \frac{1}{3}\sum_{j=1}^{3}\langle \mathcal{T}^{jj}\rangle =P_{\mathrm{kin}%
}^{n}+P_{\mathrm{kin}}^{p}  \notag \\
& -\frac{1}{2}\left[ m_{\sigma }^{2}\bar{\sigma}^{2}-m_{\omega }^{2}\bar{%
\omega}_{0}^{2}-m_{\rho }^{2}\left( \bar{\rho}_{0}^{(3)}\right) ^{2}\right]
\notag \\
& -\frac{1}{3}b_{\sigma }(g_{\sigma }\bar{\sigma})^{3}-\frac{1}{4}c_{\sigma
}(g_{\sigma }\bar{\sigma})^{4}+\frac{1}{4}c_{\omega }(g_{\omega }\bar{\omega}%
_{0})^{4}  \notag \\
& +\frac{1}{2}\left( g_{\rho }\bar{\rho}_{0}^{(3)}\right) ^{2}\left[ \Lambda
_{S}(g_{\sigma }\bar{\sigma})^{2}+\Lambda _{V}(g_{\omega }\bar{\omega}%
_{0})^{2}\right] ,
\end{align}%
where the kinetic part of pressure is given by
\begin{equation}
P_{\mathrm{kin}}^{J}=\frac{1}{3\pi ^{2}}\int_{0}^{k_{F}^{J}}dk\frac{k^{4}}{%
\sqrt{k^{2}+{M_{0}^{\ast }}^{2}}},~~J=p,n.  \label{pressureKin}
\end{equation}

The binding energy per nucleon of the asymmetric nuclear matter can be
calculated through the energy density by
\begin{equation}
E(\rho ,\delta )=\frac{\varepsilon (\rho ,\delta )}{\rho }-M.  \label{SiEn}
\end{equation}%
Furthermore, the symmetry energy $E_{\mathrm{sym}}(\rho )$ can be obtained
as
\begin{equation}
E_{\mathrm{sym}}(\rho )=\left. \frac{1}{2!}\frac{\partial ^{2}E(\rho ,\delta
)}{\partial \delta ^{2}}\right\vert _{\delta =0}=\frac{k_{F}^{2}}{%
6E_{F}^{\ast }}+\frac{g_{\rho }^{2}\rho }{2Q_{\rho }},  \label{NonRMFEsym}
\end{equation}%
while the $4$th-order symmetry energy $E_{\mathrm{{sym},4}}(\rho )$ can be
expressed as
\begin{equation}
E_{\mathrm{{sym},4}}(\rho )=\left. \frac{1}{4!}\frac{\partial ^{4}E(\rho
,\delta )}{\partial \delta ^{4}}\right\vert _{\delta =0}=E_{\mathrm{{sym},4}%
}^{\mathrm{kin}}(\rho )+E_{\mathrm{{sym},4}}^{\mathrm{M}}(\rho ),
\label{NonRMFEsym4}
\end{equation}%
where%
\begin{align}
E_{\mathrm{{sym},4}}^{\mathrm{kin}}(\rho )=& \frac{k_{F}^{2}}{648}\frac{4{%
M_{0}^{\ast 4}}+11{M_{0}^{\ast 2}}k_{F}^{2}+10k_{F}^{4}}{E_{F}^{\ast 5}},
\label{Esym4Kin} \\
E_{\mathrm{{sym},4}}^{\mathrm{M}}(\rho )=& \frac{g_{\rho }^{8}\rho ^{3}}{%
2Q_{\rho }^{4}}\left( \frac{\Lambda _{V}^{2}g_{\omega }^{4}\bar{\omega}%
_{0}^{2}}{Q_{\omega }}-\frac{\Lambda _{S}^{2}g_{\sigma }^{4}\bar{\sigma}^{2}%
}{Q_{\sigma }}\right)  \notag \\
& +\frac{g_{\sigma }^{2}\rho {M_{0}^{\ast }}k_{F}^{2}}{24Q_{\sigma
}E_{F}^{\ast 3}}\left( \frac{4\Lambda _{S}g_{\sigma }g_{\rho }^{4}\bar{\sigma%
}\rho }{Q_{\rho }^{2}}-\frac{{M_{0}^{\ast }}k_{F}^{2}}{3E_{F}^{\ast 3}}%
\right) ,  \label{Esym4M}
\end{align}%
with $E_{\mathrm{{sym},4}}^{\mathrm{kin}}(\rho )$ representing the kinetic
part (including the interactions due to the nucleon effective mass) while $%
E_{\mathrm{{sym},4}}^{\mathrm{M}}(\rho )$ the other part due to the
interaction in the $4$th-order symmetry energy, and $E_{F}^{\ast }=\sqrt{%
M_{0}^{\ast 2}+k_{F}^{2}}$. The coefficients $Q_{\sigma }$, $Q_{\omega }$
and $Q_{\rho }$ are defined as
\begin{align}
Q_{\sigma }=& m_{\sigma }^{2}+g_{\sigma }^{2}\left( \frac{3\rho _{S}}{{%
M_{0}^{\ast }}}-\frac{3\rho }{E_{F}^{\ast }}\right) +2b_{\sigma }Mg_{\sigma
}^{3}\bar{\sigma}+3c_{\sigma }g_{\sigma }^{4}\bar{\sigma}^{2},  \label{Qf} \\
Q_{\omega }=& m_{\omega }^{2}+3c_{\omega }g_{\omega }^{4}\bar{\omega}%
_{0}^{2},  \label{Qw} \\
Q_{\rho }=& m_{\rho }^{2}+\Lambda _{S}g_{\sigma }^{2}g_{\rho }^{2}\bar{\sigma%
}^{2}+\Lambda _{V}g_{\omega }^{2}g_{\rho }^{2}\bar{\omega}_{0}^{2}.
\label{Qr}
\end{align}%
In the above expressions, all the fields are calculated in the case of
symmetric nuclear matter, i.e., at $\delta =0$.

The analytical expression of the symmetry energy $E_{\mathrm{sym}}(\rho )$,
i.e., Eq. (\ref{NonRMFEsym}), is a well-known result firstly given in Ref.
\cite{Hor01}. To our best knowledge, the formulas (\ref{NonRMFEsym4})-(\ref%
{Qr}) give, for the first time, the analytical expression of the $4$th-order
symmetry energy $E_{\mathrm{{sym},4}}(\rho )$ in the RMF model, which are
the main results of the present work. These analytical expressions allow us
to evaluate accurately the $4$th-order symmetry energy $E_{\mathrm{{sym},4}%
}(\rho )$ and thus study the higher-order corrections to the empirical
parabolic approximation within the framework of the RMF model. Before
presenting numerical results, it is instructive to analyze firstly the low
density behavior of the $E_{\mathrm{{sym},4}}(\rho )$. When $\rho
\rightarrow 0$, the magnitude of all fields will approach to zero and both $%
E_{F}^{\ast }$ and $M_{0}^{\ast }$ will approach to $M$, leading to $E_{%
\mathrm{{sym},4}}^{\mathrm{M}}(\rho )\rightarrow 0$ from Eq. (\ref{Esym4M}%
) and $E_{\mathrm{{sym},4}}^{\mathrm{kin}}(\rho )\rightarrow \frac{1}{162}%
\frac{k_{F}^{2}}{M}$ from Eq. (\ref{Esym4Kin}). Therefore, in the
low density limit, we have
\begin{equation}
\lim_{\rho \rightarrow 0}E_{\mathrm{{sym},4}}(\rho )\rightarrow \frac{1}{162}%
\frac{k_{F}^{2}}{M},  \label{lowdenlimEsym4}
\end{equation}%
which is exactly the result from the free Fermi gas model as
expected.

\section{Results and Discussions}

\label{s3}

\subsection{The $4$th-order symmetry energy and higher-order effects on the
isobaric incompressibility of asymmetric nuclear matter}

\begin{figure}[tbh]
\includegraphics[scale=0.9]{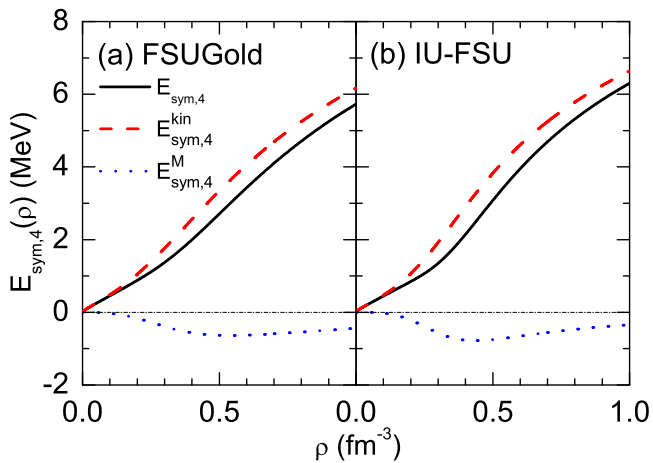}
\caption{(Color online) Density dependence of the $4$th-order symmetry
energy $E_{\mathrm{{sym},4}}(\protect\rho )$ as well as its kinetic part $E_{%
\mathrm{{sym},4}}^{\mathrm{kin}}(\protect\rho )$ and interacting part $E_{%
\mathrm{{sym},4}}^{\mathrm{M}}(\protect\rho )$ from two accurately
calibrated interactions, i.e., FSUGold (a) and IU-FSU (b).}
\label{Esym4NL3FSUIUFSU}
\end{figure}
Shown in Fig. \ref{Esym4NL3FSUIUFSU} is the density dependence of the $4$%
th-order symmetry energy $E_{\mathrm{{sym},4}}(\rho )$ as well as its
kinetic part $E_{\mathrm{{sym},4}}^{\mathrm{kin}}(\rho )$ and interacting
part $E_{\mathrm{{sym},4}}^{\mathrm{M}}(\rho )$ using two accurately
calibrated interactions, i.e., FSUGold \cite{Tod05} and IU-FSU \cite{fatt10}%
. The FSUGold has been accurately calibrated to the ground-state properties
of closed-shell nuclei, their linear response, and the structure of neutron
stars while the IU-FSU is a recently developed effective interaction that
improves the FSUGold by incorporating some of the recent constraints on
properties of neutron stars. One can see from Fig. \ref{Esym4NL3FSUIUFSU}
that the $E_{\mathrm{{sym},4}}(\rho )$ is quite small (less than $0.7$ MeV)
at normal nuclear matter density while it increases with density and can
reach to about $7$ MeV at $\rho =1$ fm$^{-3}$. Furthermore, one can see that
the kinetic part $E_{\mathrm{{sym},4}}^{\mathrm{kin}}(\rho )$ dominates over
the interacting part $E_{\mathrm{{sym},4}}^{\mathrm{M}}(\rho )$ with the
latter is generally negative.
\begin{figure}[tbh]
\includegraphics[scale=0.9]{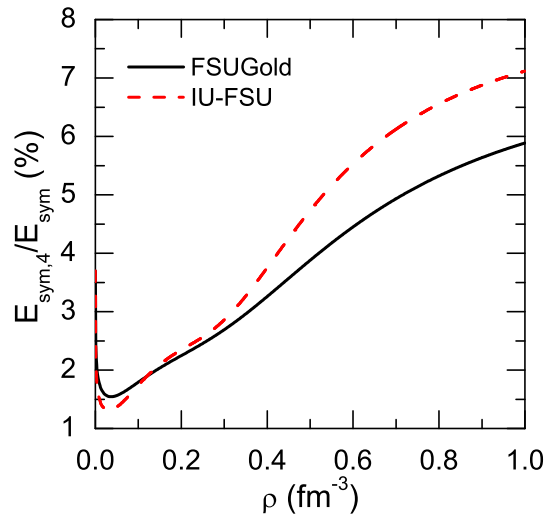}
\caption{(Color online) Ratio of the $4$th-order symmetry energy to the
symmetry energy as a function of density from two accurately calibrated
interactions, i.e., FSUGold and IU-FSU. }
\label{Esym4Esym2}
\end{figure}

In order to investigate higher-order $E_{\mathrm{{sym},4}}(\rho )$ effects
on the EOS of asymmetric nuclear matter, it may make more sense to calculate
the ratio of the $4$th-order symmetry energy to the symmetry energy, i.e., $%
E_{\mathrm{{sym},4}}(\rho )/E_{\mathrm{sym}}(\rho )$. In Fig. \ref%
{Esym4Esym2}, we show this ratio as a function of density with interactions
FSUGold and IU-FSU. It is seen that the ratio $E_{\mathrm{{sym},4}}(\rho
)/E_{\mathrm{sym}}(\rho )$ has a very small value of about $2\%$ around
normal nuclear matter density $\rho _{0}$, but it can reach to about $%
6\%\sim 7\%$ at high densities (e.g., $1.0\,\mathrm{{fm}^{-3}}$ or $6\sim
7\rho _{0}$). This result is essentially consistent with the nonrelativistic
calculations in some phenomenological models \cite{XuC11}. These features
imply that the $E_{\mathrm{{sym},4}}(\rho )$ may become important at higher
densities in some extreme physical conditions such as in neutron star where
the isospin asymmetry $\delta $ can be close to unity. As an example, in the
next subsection, we shall study effects of the $4$th-order symmetry energy
on the proton fraction in $\beta $-stable neutron star matter. In addition,
one can see from Fig. \ref{Esym4Esym2} that in the low density limit, we
have $\lim\limits_{\rho \rightarrow 0}E_{\mathrm{{sym},4}}(\rho )/E_{\mathrm{%
sym}}(\rho )=1/27$ as expected from the free Fermi gas model.
\begin{table*}[tbh]
\caption{Characteristic parameters of asymmetric nuclear matter, namely, $%
\protect\rho _{0}$ ($\mathrm{{fm}^{-3}}$), $E_{0}(\protect\rho
_{0})$ (MeV),
$E_{\mathrm{sym}}(\protect\rho _{0})$ (MeV), $E_{\mathrm{{sym},4}}(\protect%
\rho _{0})$ (MeV), $K_{0}$ (MeV), $J_{0}$ (MeV), $I_{0}$ (MeV), $L$ (MeV), $%
K_{\mathrm{sym}}$ (MeV), $J_{\mathrm{sym}}$ (MeV),
$I_{\mathrm{sym}}$ (MeV),
$L_{\mathrm{{sym},4}}$ (MeV), $K_{\mathrm{{sym},4}}$ (MeV), $J_{\mathrm{{sym}%
,4}}$ (MeV), $K_{\mathrm{asy}}$ (MeV), $K_{\mathrm{{sat},2}}$ (MeV), $K_{%
\mathrm{{sat},4}}$ (MeV), and the ratios $K_{\mathrm{{sat},2}}/K_{\mathrm{asy%
}}$ and $K_{\mathrm{{sat},4}}/K_{\mathrm{{sat},2}}$, for different interactions.}%
\begin{tabular}{|c|r|r|r|r|r|r|r|}
\hline
& FSUGold & IU-FSU & FSU-I & FSU-II & FSU-III & FSU-IV & FSU-V \\
\hline\hline
$\rho _{0}$ & $0.148$ & $0.155$ & $0.148$ & $0.148$ & $0.148$ & $0.148$ & $%
0.148$ \\ \hline $E_{0}(\rho _{0})$ & $-16.3$ & $-16.4$ & $-16.3$ &
$-16.3$ & $-16.3$ & $-16.3 $ & $-16.3$ \\ \hline
$E_{\mathrm{sym}}(\rho _{0})$ & $32.5$ & $31.3$ & $37.4$ & $35.5$ &
$33.9$ & $31.4$ & $30.9$ \\ \hline $E_{\mathrm{{sym},4}}(\rho _{0})$
& $0.66$ & $0.67$ & $0.66$ & $0.66$ & $0.66 $ & $0.66$ & $0.78$ \\
\hline $K_{0}$ & $229.2$ & $232.3$ & $229.2$ & $229.2$ & $229.2$ &
$229.2$ & $229.2$
\\ \hline
$J_{0}$ & $-521.6$ & $-288.5$ & $-521.6$ & $-521.6$ & $-521.6$ & $-521.6$ & $%
-521.6$ \\ \hline
$I_{0}$ & $2815.7$ & $4541.9$ & $2815.7$ & $2815.7$ & $2815.7$ & $2815.7$ & $%
2815.7$ \\ \hline $L$ & $60.4$ & $47.3$ & $109.5$ & $87.4$ & $71.7$
& $52.1$ & $49.4$ \\ \hline $K_{\mathrm{sym}}$ & $-51.4$ & $29.0$ &
$2.7$ & $-68.4$ & $-74.4$ & $-16.7$ & $5.5$ \\ \hline
$J_{\mathrm{sym}}$ & $426.5$ & $363.9$ & $-101.4$ & $157.6$ & $399.0$ & $%
251.2$ & $80.4$ \\ \hline $-I_{\mathrm{sym}}$ & $6331.8$ & $11346.5$
& $285.8$ & $1364.1$ & $4211.9$ & $6136.2$ & $4620.4$ \\ \hline
$L_{\mathrm{{sym},4}}$ & $1.9$ & $1.8$ & $1.9$ & $1.9$ & $1.9$ &
$1.9$ & $2.3 $ \\ \hline $K_{\mathrm{{sym},4}}$ & $0.5$ & $0.1$ &
$0.5$ & $0.5$ & $0.5$ & $0.5$ & $0.1 $ \\ \hline
$J_{\mathrm{{sym},4}}$ & $5.0$ & $6.3$ & $4.8$ & $4.8$ & $4.8$ &
$5.2$ & $5.1 $ \\ \hline
$K_{\mathrm{asy}}$ & $-413.8$ & $-255.1$ & $-654.3$ & $-592.6$ & $-504.4$ & $%
-329.4$ & $-290.9$ \\ \hline $K_{\mathrm{{sat},2}}$ & $-276.4$ &
$-196.3$ & $-405.1$ & $-393.8$ & $-341.4$ & $-210.8$ & $-178.5$ \\
\hline $K_{\mathrm{{sat},4}}$ & $3.0$ & $47.6$ & $338.6$ & $183.4$ &
$50.0$ & $12.9$ & $32.4$ \\ \hline
$K_{\mathrm{{sat},2}}/K_{\mathrm{asy}}$ & $67\%$ & $77\%$ & $62\%$ &
$66\%$ & $68\%$ & $64\%$ & $61\%$ \\ \hline
$-K_{\mathrm{{sat},4}}/K_{\mathrm{{sat},2}}$ & $1\%$ & $24\%$ & $84\%$ & $%
47\%$ & $15\%$ & $6\%$ & $18\%$ \\ \hline
\end{tabular}%
\label{TabKsat}
\end{table*}

The analytical expression of the $4$th-order symmetry energy $E_{\mathrm{{sym%
},4}}(\rho )$ allows us to calculate accurately the density slope
and curvature parameters of $E_{\mathrm{{sym},4}}(\rho )$, i.e.,
$L_{\mathrm{{sym},4}}$ and $K_{\mathrm{{sym},4}}$, and thus obtain
the accurate value of the higher-order isobaric incompressibility
$K_{\mathrm{{sat},4}}$ of asymmetric nuclear matter according to Eq.
(\ref{Ksat4}). Table \ref{TabKsat} displays the characteristic
parameters of asymmetric nuclear matter, namely, $\rho
_{0}$, $E_{0}(\rho _{0})$, $E_{\mathrm{sym}}(\rho _{0})$, $E_{\mathrm{{sym},4%
}}(\rho _{0})$, $K_{0}$, $J_{0}$, $I_{0}$, $L$, $K_{\mathrm{sym}}$, $J_{%
\mathrm{sym}}$, $I_{\mathrm{sym}}$, $L_{\mathrm{{sym},4}}$, $K_{\mathrm{{sym}%
,4}}$, $J_{\mathrm{{sym},4}}$, $K_{\mathrm{asy}}$, $K_{\mathrm{{sat},2}}$, $%
K_{\mathrm{{sat},4}}$ and the ratios $K_{\mathrm{{sat},2}}/K_{\mathrm{asy}}$
and $K_{\mathrm{{sat},4}}/K_{\mathrm{{sat},2}}$, for the two accurately
calibrated interactions FSUGold and IU-FSU. To see the variation of the
higher-order characteristic parameters with the density dependence of the
symmetry energy, we also include in Table \ref{TabKsat} the results from $5$
interactions denoted as FSU-I, FSU-II, FSU-III, FSU-IV and FSU-V for which
the parameters $(\Lambda _{S},\Lambda _{V})$ are selected as $(0.00,0.00)$, $%
(0.00,0.01)$, $(0.00,0.02)$, $(0.00,0.04)$ and $(0.01,0.03)$, respectively,
while the $g_{\rho }$ parameter is adjusted accordingly to fix $E_{\mathrm{%
sym}}(\rho _{f})=25.57\, \mathrm{MeV}$ at $\rho _{f}=0.1\,\mathrm{{fm}^{-3}}$
as in the FSUGold interaction. The other parameters for FSU-I, FSU-II,
FSU-III, FSU-IV and FSU-V are exactly the same as in FSUGold (Note: the
FSUGold corresponds to the case of $(\Lambda _{S},\Lambda _{V})=(0.00,0.03)$%
). This is equivalent to solve a constraint equation about $g_{\rho
},\Lambda _{S}$ and $\Lambda _{V}$, i.e., $\frac{g_{\rho }^{2}\rho }{%
2Q_{\rho }}=13.47\,\mathrm{MeV}$ where $\bar{\sigma}$ and
$\bar{\omega}_{0}$ in $Q_{\rho }$ (See Eq. ((\ref{Qr}))) are
determined by the properties of symmetric nuclear matter in FSUGold.

From Table \ref{TabKsat}, one can see that the
$E_{\mathrm{{sym},4}}(\rho _{0})$ is generally less than $1$ MeV
(about $0.66$ MeV for most of the
interactions considered here), consistent with that observed in Fig. \ref%
{Esym4NL3FSUIUFSU}. These results about the $E_{\mathrm{{sym},4}}(\rho _{0})$
are further in agreement with the calculations in the nonrelativistic models
of MDI and Skyrme-Hartree-Fock \cite{Che09}, nicely verifying the empirical
parabolic law around the normal nuclear matter density $\rho _{0}$.

As pointed out previously, the difference between $K_{\mathrm{{sat},2}}$ and
$K_{\mathrm{asy}}$ reflects the contribution from higher-order effects,
namely, the value of ${J_{0}L}/{K_{0}}$, which has been usually neglected in
many calculations in the literature \cite{baran05,Che05,Cen09,LiT07,lopez88}%
. From Table \ref{TabKsat}, one can see that neglecting the ${J_{0}L}/{K_{0}}
$ term generally leads to $30$-$40\%$ relative error for the $K_{\mathrm{{sat%
},2}}$ parameter and thus the ${J_{0}L}/{K_{0}}$ term contribution to the $%
K_{\mathrm{{sat},2}}$ parameter cannot be neglected simply, confirming the
previous findings in the nonrelativistic studies \cite{Che09}.

Furthermore, it is seen from Table \ref{TabKsat} that the value of
higher-order $K_{\mathrm{{sat},4}}$ is generally small compared with that of
$K_{\mathrm{{sat},2}}$ for most of the interactions considered here. In
addition, one can see that the $K_{\mathrm{{sat},4}}$ becomes more important
for the interactions with larger $L$ values and this is consistent with the
nonrelativistic studies \cite{Che09}. It should be noted that the
higher-order $K_{\mathrm{{sat},4}}$ term can be safely neglected in the
study of giant resonance of finite nuclei \cite{Pie09} where the isospin
asymmetry $\delta $ is usually small, i.e., about $0.2$.

\subsection{Effects of $E_{\mathrm{{sym},4}}$ on the proton fraction in $%
\protect\beta $-stable nuclear matter}

In order to further illustrate the effects of the $4$th-order symmetry
energy on the EOS of asymmetric nuclear matter, we calculate the proton
fraction $x_{p}$ in $\beta $-stable neutron star matter where the isospin
asymmetry $\delta $ is generally close to $1$. The chemical composition of
the neutron star is determined by the requirement of charge neutrality and
equilibrium with respect to the weak interaction ($\beta $-stable matter).
From the binding energy per nucleon, i.e., Eq. (\ref{EOSANM}), we can
calculate the proton fraction, $x_{p}=(1-\delta )/2$, for $\beta $-stable
nuclear matter as found in interior of neutron stars. For neutrino free $%
\beta $-stable nuclear matter, the chemical equilibrium for the reactions $%
n\rightarrow p+e^{-}+\bar{\nu}_{e}$ and $p+e^{-}\rightarrow n+\nu _{e}$
requires
\begin{eqnarray}
\mu _{e} &=&\mu _{n}-\mu _{p}=2\frac{\partial E}{\partial \delta }  \notag \\
&=&4\delta E_{\mathrm{sym}}(\rho )+8\delta ^{3}E_{\mathrm{{sym},4}}(\rho )+%
\mathcal{O}(\delta ^{5})  \label{elechem1}
\end{eqnarray}%
where $\mu _{i}=\partial E_{i}/\partial x_{i}~(i=n,p,e,\mu )$ is the
chemical potential. For relativistic degenerate electrons, we have
\begin{eqnarray}
\mu _{e} &=&\left( m_{e}^{2}+k_{F}^{e2}\right) ^{1/2}  \notag \\
&=&\left[ m_{e}^{2}+(3\pi ^{2}\rho x_{e})^{2/3}\right] ^{1/2}\simeq \left(
3\pi ^{2}\rho x_{e}\right) ^{1/3}  \label{elechem2}
\end{eqnarray}%
where $m_{e}=0.511\,$MeV is the electron mass, and $x_{p}=x_{e}$ because of
charge neutrality.

Just above a nuclear matter density at which $\mu _{e}$ exceeds the muon
mass $m_{\mu }=0.105\,$GeV, the reactions $e^{-}\rightarrow \mu ^{-}+\nu
_{e}+\bar{\nu}_{\mu }$, $p+\mu ^{-}\rightarrow n+\nu _{\mu }$ and $%
n\rightarrow p+\mu ^{-}+\bar{\nu}_{\mu }$ are energetically allowed so that
both electrons and muons are present in $\beta $-stable nuclear matter, this
alters $\beta $-stability condition to
\begin{equation}
\mu _{n}-\mu _{p}=\mu _{e},~~\mu _{n}-\mu _{p}=\mu _{\mu }=\left[ m_{\mu
}^{2}+(3\pi ^{2}\rho x_{\mu })^{2/3}\right] ^{1/2}  \label{elemuochem}
\end{equation}%
with $x_{p}=x_{e}+x_{\mu }$.
\begin{figure}[tbh]
\includegraphics[scale=0.86]{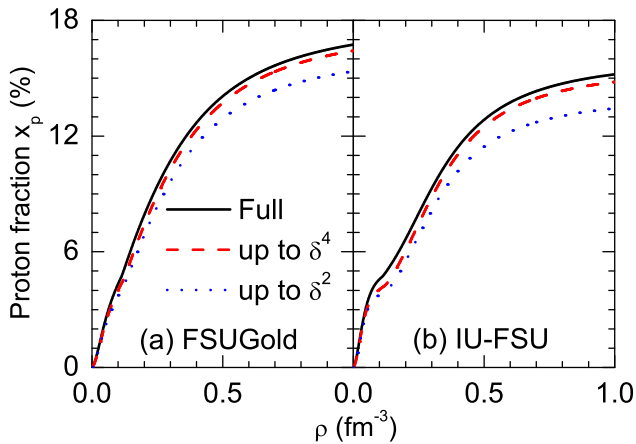}
\caption{(Color online) Density dependence of the proton fraction $x_{p}$ in
$\protect\beta $-stable $npe\protect\mu $ matter with FSUGold (a) and IU-FSU
(b). Three cases, i.e., the full EOS of asymmetric nuclear matter (solid
lines), its parabolic approximation (up to $\protect\delta ^{2}$ in Eq. (%
\protect\ref{EOSANM})) (dotted lines), and further including the $4$th-order
symmetry energy (up to $\protect\delta ^{4}$ in Eq. (\protect\ref{EOSANM}))
(dashed lines), are considered.}
\label{Xproton}
\end{figure}

In Fig. \ref{Xproton}, we show the density dependence of the proton fraction
$x_{p}$ in $\beta $-stable $npe\mu $ matter with the interactions FSUGold
and IU-FSU. We consider three cases for the EOS of asymmetric nuclear matter
here, i.e., the full one, its parabolic approximation (up to $\delta ^{2}$
in Eq. (\ref{EOSANM})), and\ the one further including the $4$th-order
symmetry energy (up to $\delta ^{4}$ in Eq. (\ref{EOSANM})). The results
show that, for both interactions of FSUGold and IU-FSU, the $4$th-order
symmetry energy is moderately important for the proton fraction, especially
at higher densities. For the FSUGold (IU-FSU)\ interaction and $\rho =1.0\,%
\mathrm{{fm}^{-3}}$, for instance, including the $4$th-order symmetry energy
in the parabolic approximation to the EOS of asymmetric nuclear matter will
increase the proton fraction $x_{p}$ from $15.37\%$ ($13.44\%$) to $16.43\%$
($14.81\%$), producing a relative variation of about $7\%$ ($10\%$). These
results indicate that the $4$th-order symmetry energy may have obvious
effects on the proton fraction $x_{p}$ in $\beta $-stable $npe\mu $ matter
and the parabolic approximation to the EOS of asymmetric nuclear matter may
significantly underestimate the proton fraction, especially at higher
densities. These features are consistent with the nonrelativistic
Skyrme-Hartree-Fock calculations \cite{Zha01}.

Furthermore, one can see from Fig. \ref{Xproton} that the difference between
the results with the full EOS and with the one containing the terms up to
the $4$th-order symmetry energy is very small, indicating that the EOS of
asymmetric nuclear matter including the terms up to the $4$th-order symmetry
energy (up to $\delta ^{4}$ in Eq. (\ref{EOSANM})) could be a good
approximation for the determination of the proton fraction in $\beta $%
-stable $npe\mu $ matter.

\subsection{Effects of $E_{\mathrm{{sym},4}}$ on core-crust transition
density and pressure in neutron stars}

The transition density $\rho _{t}$ is the baryon number density that
separates the liquid core from the inner crust in neutron stars and it plays
an important role in determining many properties of neutron stars \cite%
{Hor01,Pro06,Duc08a,Duc08b,Lat07,xu09}. One simple and widely used way to
determine the core-crust transition density $\rho _{t}$ is the so-called
thermodynamical method, which requires the system to obey the following
intrinsic stability condition~\cite{Cal85,Kub07,Lat07}
\begin{eqnarray}
-\left( \frac{\partial P}{\partial v}\right) _{\mu _{np}} &>&0,
\label{ther1} \\
-\left( \frac{\partial \mu _{np}}{\partial q_{c}}\right) _{v} &>&0,
\label{ther2}
\end{eqnarray}%
where the $P=P_{b}+P_{e}$ is the total pressure of the $npe$ matter system
with $P_{b}$ and $P_{e}$ denoting the contributions from baryons and
electrons respectively, and the $v$ and $q_{c}$ are the volume and charge
per baryon number. The $\mu _{np}$ is defined as the chemical potential
difference between neutrons and protons, i.e., $\mu _{np}=\mu _{n}-\mu _{p}$%
. The pressure $P_{e}$ is only a function of the chemical potential
difference $\mu _{np}$ by assuming the $\beta $-equilibrium condition is
satisfied, i.e., $\mu _{np}=\mu _{e}$. By using the relation ${\partial
E_{b}(\rho ,x_{p})}/{\partial x_{p}}=-\mu _{np}$ with $E_{b}(\rho ,x_{p})$
being energy per baryon from the baryons in the $\beta $-equilibrium neutron
star matter and $x_{p}=\rho _{p}/\rho $, and treating the electrons as free
Fermi gas, one can show \cite{xu09} that the thermodynamical relations Eq.~(%
\ref{ther1}) and Eq.~(\ref{ther2}) are actually equivalent to the following
condition

\begin{widetext}

\begin{eqnarray}
V_{\mathrm{thermal}} &=&2\rho \frac{\partial E_{b}(\rho
,x_{p})}{\partial \rho }+\rho ^{2}\frac{\partial ^{2}E_{b}(\rho
,x_{p})}{\partial \rho ^{2}} -\left. \left( \frac{\partial
^{2}E_{b}(\rho ,x_{p})}{\partial \rho
\partial x_{p}}\rho \right) ^{2}\right/ \frac{\partial ^{2}E_{b}(\rho ,x_{p})%
}{\partial x_{p}^{2}}>0,  \label{Vther}
\end{eqnarray}%
which determines the thermodynamical instability region of the $\beta $%
-equilibrium neutron star matter. The baryon number density that violates
the condition Eq.~(\ref{Vther}) then corresponds to the core-crust
transition density in neutron stars for the thermodynamical method.

With the EOS of asymmetric nuclear matter including the terms up to the $4$%
th-order symmetry energy (i.e., up to $\delta ^{4}$ in Eq. (\ref{EOSANM})),
Eq.~(\ref{Vther}) is then reduced to

\begin{eqnarray}
V_{\mathrm{thermal}} &=&\rho ^{2}\frac{\partial ^{2}E_{0}}{\partial \rho ^{2}%
}+2\rho \frac{\partial E_{0}}{\partial \rho }+\delta ^{2}\left[ \rho ^{2}%
\frac{\partial ^{2}E_{\mathrm{sym}}(\rho )}{\partial \rho ^{2}}+2\rho \frac{%
\partial E_{\mathrm{sym}}(\rho )}{\partial \rho }\right] +\delta ^{4}\left[
\rho ^{2}\frac{\partial ^{2}E_{\mathrm{sym,4}}(\rho )}{\partial \rho ^{2}}%
+2\rho \frac{\partial E_{\mathrm{sym,4}}(\rho )}{\partial \rho }\right]
\notag \\
&&-\frac{\rho ^{2}\delta ^{2}}{E_{\mathrm{sym}}(\rho )+6E_{\mathrm{sym,4}%
}(\rho )}\left[ \frac{\partial E_{\mathrm{sym}}(\rho )}{\partial \rho }%
+2\delta ^{2}\frac{\partial E_{\mathrm{sym,4}}(\rho )}{\partial \rho }\right]
^{2}>0.  \label{Vther4th}
\end{eqnarray}%
The baryon number density that violates the condition Eq.~(\ref%
{Vther4th}) then corresponds to the core-crust transition density
$\rho _{t}^{\mathrm{4th}}$ in neutron stars for the EOS of
asymmetric nuclear
matter including the terms up to the $4$th-order symmetry energy (up to $%
\delta ^{4}$ in Eq. (\ref{EOSANM})). The corresponding transition pressure $%
P_{t}^{\mathrm{4th}}$ at $\rho _{t}^{\mathrm{4th}}$ for the EOS of
asymmetric nuclear matter including the terms up to the $4$th-order
symmetry energy (up
to $\delta ^{4}$ in Eq. (\ref{EOSANM})) is then given by%
\begin{eqnarray}
P_{t}^{\mathrm{4th}}
&=&P_{t}^{b,\mathrm{4th}}+P_{t}^{e,\mathrm{4th}} \notag
\\
&=&\left[ \rho ^{2}\left( \frac{\partial E_{0}}{\partial \rho }+\delta ^{2}%
\frac{\partial E_{\mathrm{sym}}}{\partial \rho }+\delta ^{4}\frac{\partial
E_{\mathrm{sym,4}}}{\partial \rho }\right) \right] _{\rho =\rho _{t}}+\mu
_{e}^{\mathrm{4th}}\rho _{e}  \notag \\
&=&\left[ \rho ^{2}\left( \frac{\partial E_{0}}{\partial \rho }+\delta ^{2}%
\frac{\partial E_{\mathrm{sym}}}{\partial \rho }+\delta ^{4}\frac{\partial
E_{\mathrm{sym,4}}}{\partial \rho }\right) \right] _{\rho =\rho
_{t}}+\left\{ 4\rho \delta \cdot \frac{1-\delta }{2}\cdot \lbrack E_{\mathrm{%
sym}}+\delta ^{2}E_{\mathrm{sym,4}}]\right\} _{\rho =\rho _{t},\delta
=\delta _{t}}
\end{eqnarray}%
where $\delta _{t}$ is the isospin asymmetry of the $\beta $%
-equilibrium neutron star matter at the corresponding transition density.

The transition density obtained by neglecting the $E_{\mathrm{sym,4}}(\rho )$
term in the condition Eq.~(\ref{Vther4th}) corresponds to the core-crust
transition density $\rho _{t}^{\mathrm{2nd}}$ in neutron stars for the
parabolic approximation to the EOS of asymmetric nuclear matter (up to $%
\delta ^{2}$ in Eq. (\ref{EOSANM})). The corresponding transition
pressure at $P_{t}^{\mathrm{2nd}}$ for the parabolic approximation
to the EOS of asymmetric nuclear matter (up to $\delta ^{2}$ in Eq.
(\ref{EOSANM})) is then
expressed as%
\begin{eqnarray}
P_{t}^{\mathrm{2nd}}
&=&P_{t}^{b,\mathrm{2nd}}+P_{t}^{e,\mathrm{2nd}} \notag
\\
&=&\left[ \rho ^{2}\left( \frac{\partial E_{0}}{\partial \rho }+\delta ^{2}%
\frac{\partial E_{\mathrm{sym}}}{\partial \rho }\right) \right] _{\rho =\rho
_{t}}+\mu _{e}^{\mathrm{2nd}}\rho _{e}=\left[ \rho ^{2}\left( \frac{\partial
E_{0}}{\partial \rho }+\delta ^{2}\frac{\partial E_{\mathrm{sym}}}{\partial
\rho }\right) \right] _{\rho =\rho _{t}}+\left[ 4\rho \delta \cdot \frac{%
1-\delta }{2}\cdot E_{\mathrm{sym}}\right] _{\rho =\rho _{t},\delta =\delta
_{t}}.
\end{eqnarray}%
\end{widetext}
Due to simplicity, the $\rho _{t}^{\mathrm{2nd}}$ and $P_{t}^{\mathrm{2nd}}$
have been extensively applied to determine the inner edge of neutron star
crusts within the nonrelativistic models \cite{Kub07,Lat07,Oya07,Wor08} and
recently in the RMF model \cite{Mou10} as well. However, recent studies
based on some nonrelativistic models have demonstrated \cite{xu09} that the
parabolic approximation to the EOS of asymmetric nuclear matter may lead
systematically to significantly higher core-crust transition densities and
pressures, especially with stiffer symmetry energy functionals. It is thus
very interesting to see how the higher-order $E_{\mathrm{sym,4}}(\rho )$
affects the transition density $\rho _{t}$ and pressure $P_{t}$ in the RMF
model.

In Table \ref{rhoT}, we show the $\rho _{t}^{\mathrm{2nd}}$, $\rho _{t}^{%
\mathrm{4th}}$, $P_{t}^{\mathrm{2nd}}$, and $P_{t}^{\mathrm{4th}}$ obtained
from the thermodynamical method with different interactions as in In Table %
\ref{TabKsat}. It is interesting to see that including the $4$th-order
symmetry energy in the parabolic approximation to the EOS of asymmetric
nuclear matter indeed reduces significantly the core-crust transition
density $\rho _{t}$, which is consistent with the nonrelativistic
calculations \cite{xu09}. Furthermore, one can see that the $4$th-order
symmetry energy may have even more drastic effects on the core-crust
transition pressure $P_{t}$, namely, including the $4$th-order symmetry
energy in the parabolic approximation to the EOS of asymmetric nuclear
matter reduces drastically the core-crust transition pressure $P_{t}$.
Therefore, our results indicate that the empirical parabolic approximation
may cause large errors for the determination of the $\rho _{t}$ and $P_{t}$
in neutron stars in the nonlinear RMF model. 

\begin{table*}[tbh]
\caption{The $\protect\rho _{t}^{\mathrm{2nd}}$ (fm$^{-3}$), $\protect\rho %
_{t}^{\mathrm{4th}}$ (fm$^{-3}$), $P_{t}^{\mathrm{2nd}}$ (MeV/fm$^{3}$), and
$P_{t}^{\mathrm{4th}}$ (MeV/fm$^{3}$) obtained from the thermodynamical
method with different interactions.}
\label{rhoT}\centering%
\begin{tabular}{|c|r|r|r|r|r|r|r|}
\hline
& FSUGold & IU-FSU & FSU-I & FSU-II & FSU-III & FSU-IV & FSU-V \\
\hline\hline
$\rho _{t}^{\mathrm{2nd}}$ & $0.089$ & $0.090$ & $0.085$ & $0.088$ & $0.088$
& 0.083 & 0.080 \\ \hline
$\rho _{t}^{\mathrm{4th}}$ & $0.051$ & $0.077$ & $0.069$ & $0.054$ & $0.053$
& $0.068$ & 0.072 \\ \hline
$P_{t}^{\mathrm{2nd}}$ & $1.316$ & $0.673$ & $0.664$ & $1.010$ & $0.968$ &
0.621 & 0.501 \\ \hline
$P_{t}^{\mathrm{4th}}$ & $0.321$ & $0.530$ & $0.302$ & $0.236$ & $0.259$ &
0.420 & 0.414 \\ \hline
\end{tabular}%
\end{table*}


\section{Summary}

\label{s4}

We have derived for the first time the analytical expression of the
nuclear matter fourth-order symmetry energy
$E_{\mathrm{{sym},4}}(\rho )$ within the framework of the nonlinear
RMF model. It should be mentioned that the analytical expression of
$E_{\mathrm{{sym},4}}(\rho )$ can be easily generalized to the case
of the density dependent RMF model that has similar isospin
structure as the nonlinear RMF model (See, e.g., Ref.~\cite{Che07}).
This provides the possibility to investigate the higher-order
$E_{\mathrm{{sym},4}}(\rho )$ corrections to the widely used
empirical parabolic law for the asymmetric nuclear matter in the RMF
model. In the present work, as examples, we have investigated the
$E_{\mathrm{{sym},4}}(\rho )$ effects on the properties of
asymmetric nuclear matter, the proton fraction $x_{p}$ in $\beta $-stable $%
npe\mu $ matter and the core-crust transition density $\rho _{t}$ and
pressure $P_{t}$ in neutron stars within the nonlinear RMF model with two
accurately calibrated interactions, i.e., FSUGold and IU-FSU.

Firstly, our results have indicated that the value of $E_{\mathrm{{sym},4}%
}(\rho )$ at normal nuclear matter density $\rho _{0}$ is generally less
than $1$ MeV, and thus the empirical parabolic approximation $E(\rho ,\delta
)\simeq E_{0}(\rho )+E_{\mathrm{sym}}(\rho )\delta ^{2}$ has been nicely
confirmed around $\rho _{0}$. However, at higher densities such as $1$ fm$%
^{-3}$, the value of $E_{\mathrm{{sym},4}}(\rho )$ can be about $7$ MeV and
the ratio of $E_{\mathrm{{sym},4}}(\rho )/E_{\mathrm{sym}}(\rho )$ can reach
to about $7\%$. These results imply that the $E_{\mathrm{{sym},4}}(\rho )$
may become nonnegligible at higher densities. Furthermore, the analytical
form of the $E_{\mathrm{{sym},4}}(\rho )$ allows us to study the
higher-order effects on the isobaric incompressibility of asymmetric nuclear
matter. Our results have indicated that the value of higher-order $K_{%
\mathrm{{sat},4}}$ is generally small compared with that of $K_{\mathrm{{sat}%
,2}}$, confirming the previous nonrelativistic calculations \cite{Che09}.

Secondly, for the proton fraction $x_{p}$ in $\beta $-stable $npe\mu $
matter, we have found that, compared with the results from the empirical
parabolic approximation to the EOS of asymmetric nuclear matter, including
the $4$th-order symmetry energy $E_{\mathrm{{sym},4}}(\rho )$ can enhance
the proton fraction $x_{p}$ by about $10\%$ at higher densities. These
results indicate that the empirical parabolic approximation to the EOS of
asymmetric nuclear matter may cause obvious errors for the determination of
the proton fraction in neutron stars within the nonlinear RMF model, which
is in agreement with the results from the nonrelativistic models \cite{Zha01}%
.

Finally, we have demonstrated that including the $4$th-order symmetry energy
$E_{\mathrm{{sym},4}}(\rho )$ in the parabolic approximation to the EOS of
asymmetric nuclear matter can reduce significantly the core-crust transition
density $\rho _{t}$ and furthermore it has even more drastic effects on the
core-crust transition pressure $P_{t}$. Therefore, our results have clearly
demonstrated that the extensively used empirical parabolic approximation to
the EOS of asymmetric nuclear matter may lead systematically to
significantly higher core-crust transition density $\rho _{t}$ and pressure $%
P_{t}$ in neutron stars within the nonlinear relativistic mean field model,
confirming the previous finding based on nonrelativistic calculations \cite%
{xu09}.

Therefore, we conclude that the higher-order
$E_{\mathrm{{sym},4}}(\rho )$ in the EOS of asymmetric nuclear
matter may have different effects on different quantities, and
generally one cannot simply neglect them, especially under some
extreme physical conditions, such as in neutron stars.

\section*{ACKNOWLEDGMENTS}

This work was supported in part by the National Natural Science
Foundation of China under Grant Nos. 10975097 and 11135011, the
Shanghai Rising-Star Program under grant No. 11QH1401100, ``Shu
Guang" project supported by Shanghai Municipal Education Commission
and Shanghai Education Development Foundation, the Program for
Professor of Special Appointment (Eastern Scholar) at Shanghai
Institutions of Higher Learning, and the National Basic Research
Program of China (973 Program) under Contract No. 2007CB815004.


\begin{thebibliography}{99}
\bibitem{LiBA98} B.A. Li, C.M. Ko, and W. Bauer, topical review, Int. Jour.
Mod. Phys. E \textbf{7}, 147 (1998).

\bibitem{danie02} P. Danielewicz, R. Lacey, and W.G. Lynch, Science \textbf{%
298}, 1592 (2002).

\bibitem{lattimer04} J.M. Lattimer and M. Prakash, Science \textbf{304}, 536
(2004).

\bibitem{baran05} V. Baran, M. Colonna, V. Greco, and M. Di Toro, Phys. Rep.
\textbf{410}, 335 (2005).

\bibitem{steiner05} A.W. Steiner, M. Prakash, J.M. Lattimer, and P.J. Ellis,
Phys. Rep. \textbf{411}, 325 (2005).

\bibitem{chen07} L.W. Chen, C.M. Ko, B.A. Li, and G.C. Yong, Front. Phys.
China \textbf{2}, 327 (2007) [arXiv:0704.2340].

\bibitem{li08} B.A. Li, L.W. Chen, and C.M. Ko, Phys. Rep. \textbf{464}, 113
(2008).

\bibitem{youngblood99} D.H. Youngblood, H.L. Clark, and Y.-W. Lui, Phys.
Rev. Lett. \textbf{82}, 691 (1999).

\bibitem{LiT07} T. Li et al., Phys. Rev. Lett. \textbf{99}, 162503 (2007);
U. Garg et al., Nucl. Phys. \textbf{A788}, 36c (2007); T. Li et al.,
Phys. Rev. C \textbf{81}, 034309 (2010).

\bibitem{aic85} J. Aichelin and C. M. Ko, Phys. Rev. Lett. \textbf{55}, 2661
(1985).

\bibitem{fuchs06} C. Fuchs, Prog. Part. Nucl. Phys. \textbf{56}, 1 (2006).

\bibitem{Che05} L.W. Chen, C.M. Ko, and B.A. Li, Phys. Rev. Lett. \textbf{94}%
, 032701 (2005); Phys. Rev. C \textbf{72}, 064309 (2005); B.A. Li and L.W.
Chen, Phys. Rev. C \textbf{72}, 064611 (2005).

\bibitem{Tsa09} M.B. Tsang, Y. Zhang, P. Danielewicz, M. Famiano, Z. Li, W.
G. Lynch, and A. W. Steiner, Phys. Rev. Lett. \textbf{102}, 122701 (2009).

\bibitem{Cen09} M. Centelles, X. Roca-Maza, X. Vi\~{n}as, and M. Warda,
Phys. Rev. Lett \textbf{102}, 122502 (2009); M. Warda, X. Vi\~{n}as, X.
Roca-Maza, and M. Centelles, Phys. Rev. C \textbf{80}, 024316 (2009).

\bibitem{Nat10} J.B. Natowitz, G. R\"{o}pke, S. Typel, D. Blaschke, A.
Bonasera, K. Hagel, T. Kl\"{a}hn, S. Kowalski, L. Qin, S. Shlomo, R. Wada,
and H. H. Wolter, Phys. Rev. Lett \textbf{104}, 202501 (2010).

\bibitem{XuC10} C. Xu, B.A. Li, and L.W. Chen, Phys. Rev. C \textbf{82},
054607 (2010).

\bibitem{Che10} L.W. Chen, C.M. Ko, B.A. Li, and J. Xu, Phys. Rev. C \textbf{%
82}, 024321 (2010).

\bibitem{Tsa11} M.B. Tsang, Z. Chajecki, D. Coupland, P. Danielewicz, F.
Famiano, R. Hodges, M. Kilburn, F. Lu, W.G. Lynch, J. Winkelbauer, M.
Youngs, Y.X. Zhang, Prog. Part. Nucl. Phys. \textbf{66}, 400 (2011).

\bibitem{Che11a} L.W. Chen, Phys. Rev. C \textbf{83}, 044308 (2011).

\bibitem{New11} W.G. Newton, M. Gearheart, J. Hooker, and B.A. Li,
arXiv:1112.2018.

\bibitem{Xia09} Z.G. Xiao, B.A. Li, L.W. Chen, G.C. Yong, and M. Zhang,
Phys. Rev. Lett. \textbf{102}, 062502 (2009).

\bibitem{Fen10} Z.Q. Feng and G.M. Jin, Phys. Lett. \textbf{B683}, 140
(2010).

\bibitem{Rus11} P. Russotto \textit{et al.}, Phys. Lett. \textbf{B697}, 471
(2011).

\bibitem{XuC10b} C. Xu and B.A. Li, Phys. Rev. C \textbf{81}, 064612 (2010).

\bibitem{Zha01} F.S. Zhang and L.W. Chen, Chin. Phys. Lett. \textbf{18}, 142
(2001).

\bibitem{steiner08} A.W. Steiner, Phys. Rev. C \textbf{74}, 045808 (2006).

\bibitem{xu09} J. Xu, L.W. Chen, B.A. Li, and H.R. Ma, Phys. Rev. C \textbf{%
79}, 035802 (2009); Astrophys. J. \textbf{697}, 1549 (2009).

\bibitem{Che09} L.W. Chen, B.J. Cai, C.M. Ko, B.A. Li, C. Shen and J. Xu,
Phys. Rev. C \textbf{80}, 014322 (2009).

\bibitem{Wal74} J.D.Walecka, Ann. Phys. (NY) \textbf{83}, 491 (1974).

\bibitem{Ser86} B.D. Serot and J.D. Walecka, Adv. Nucl. Phys. \textbf{16}, 1
(1986); Int. J. Mod. Phys. E \textbf{6}, 515 (1997).

\bibitem{Men06} J. Meng, H. Toki, S.G. Zhou, S.Q. Zhang, W.H. Long, L.S.
Geng, Prog. Part. Nucl. Phys. \textbf{57}, 470 (2006).

\bibitem{sie70} P.J. Siemens, Nucl. Phys. \textbf{A141}, 225 (1970).

\bibitem{sj74} O. Sj$\ddot{\mathrm{o}}$berg, Nucl. Phys. \textbf{A222}, 161
(1974).

\bibitem{lag81} I.E. Lagaris and V.R. Pandharipande, Nucl. Phys. \textbf{A369%
}, 470 (1981).

\bibitem{prak85} M. Prakash and K. S. Bedell, Phys. Rev. C \textbf{32}, 1118
(1985).

\bibitem{lopez88} M. Lopez-Quelle, S. Marcos, R. Niembro, A. Bouyssy, and N.
V. Giai, Nucl. Phys. \textbf{A483}, 479 (1988).

\bibitem{Mul96} H. M\"{u}ller and B. D. Serot, Nucl. Phys. \textbf{A606},
508 (1996).

\bibitem{Hor01} C.J. Horowitz, and J. Piekarewicz, Phys. Rev. Lett \textbf{86%
}, 5647 (2001); Phys. Rev. C \textbf{64}, 062802(R) (2001); Phys. Rev. C
\textbf{66}, 055803 (2002).

\bibitem{Tod05} B.G. Todd-Rutel and J. Piekarewicz, Phys. Rev. Lett. \textbf{%
95}, 122501 (2005).

\bibitem{fatt10} F.J. Fattoyev, C.J. Horowitz, J. Piekarewicz, and G. Shen,
Phys. Rev. C \textbf{82}, 055803 (2010).

\bibitem{XuC11} C. Xu, B.A. Li, L.W. Chen, and C.M. Ko, Nucl. Phys. \textbf{%
A865}, 1 (2011).

\bibitem{Pie09} J. Piekarewicz and M. Centelles, Phys. Rev. C \textbf{79},
054311 (2009).

\bibitem{Pro06} C. Provid\^{e}ncia, L. Brito, S.S. Avancini, D.P. Menezes,
and P. Chomaz, Phys. Rev. C \textbf{73}, 025805 (2006).

\bibitem{Duc08a} C. Ducoin, J. Margueron, and P. Chomaz, Nucl. Phys. \textbf{%
A809}, 30 (2008).

\bibitem{Duc08b} C. Ducoin, C. Provid\^{e}ncia, A.M. Santos, L. Brito, and
P. Chomaz, Phys. Rev. C \textbf{78}, 055801 (2008).

\bibitem{Lat07} J.M. Lattimer and M. Prakash, Phys. Rep. \textbf{442}, 109
(2007).

\bibitem{Cal85} H.B. Callen, Thermodynamics, Wiley, New York, 1985.

\bibitem{Kub07} S. Kubis, Phys. Rev. C \textbf{76}, 025801 (2007); Phys.
Rev. C \textbf{70}, 065804 (2004).

\bibitem{Oya07} K. Oyamatsu and K. Iida, Phys. Rev. C \textbf{75}, 015801
(2007).

\bibitem{Wor08} A. Worley, P.G. Krastev, and B.A. Li, Astrophys. J. \textbf{%
685}, 390 (2008).

\bibitem{Mou10} Ch.C. Moustakidis, T. Nik\v{s}i\'{c}, G.A. Lalazissis, D.
Vretenar, and P. Ring, Phys. Rev. C \textbf{81}, 065803 (2010).

\bibitem{Che07} L.W. Chen, C.M. Ko, and B.A. Li, Phys. Rev. C \textbf{76},
054316 (2007).
\end{thebibliography}
\end{document}